\documentclass[aps,prl,twocolumn,superscriptaddress,showpacs,preprintnumbers,floatfix,amsmath,amssymb]{revtex4}
\pdfoutput=1

\usepackage{graphicx}
\usepackage{subfigure}
\usepackage{isotope}
\usepackage{hyperref}
\usepackage{color}

\newcommand{\GENIE}{\textsc{genie}}
\newcommand{\GLoBES}{\textsc{gl}{\small o}\textsc{bes}}

\newcommand{\ie}{\textit{i.e.}, }

\begin{document}

\preprint{FERMILAB-PUB-15-320-T}
\preprint{RM3-TH/15-12}

\title{Missing energy and the measurement of the CP-violating phase in neutrino oscillations}

\author{A. M. Ankowski}
\affiliation{Center for Neutrino Physics, Virginia Tech, Blacksburg, Virginia 24061, USA}
\author{P. Coloma}
\email{pcoloma@fnal.gov}
\affiliation{Fermi National Accelerator Laboratory, Batavia, Illinois 60510, USA}
\author{P. Huber}
\affiliation{Center for Neutrino Physics, Virginia Tech, Blacksburg, Virginia 24061, USA}
\author{C. Mariani}
\affiliation{Center for Neutrino Physics, Virginia Tech, Blacksburg, Virginia 24061, USA}
\author{E. Vagnoni}
\affiliation{INFN and Dipartimento di Matematica e Fisica, Universit\`a di Roma Tre, Via della Vasca Navale 84, 00146 Rome, Italy}

\date{\today}%

\begin{abstract}
In the next generation of long-baseline neutrino oscillation experiments, aiming to determine the charge-parity violating phase $\delta_{CP}$ in the appearance channel, fine-grained time-projection chambers are expected to play an important role. In this Letter, we analyze an influence of realistic detector capabilities on the $\delta_{CP}$ sensitivity for a setup similar to that of the Deep Underground Neutrino Experiment. We find that the effect of the missing energy, carried out by undetected particles, is sizable. Although the reconstructed neutrino energy can be corrected for the missing energy, the accuracy of such procedure has to exceed 20\%, to avoid a sizable bias in the extracted $\delta_{CP}$ value.
\end{abstract}



\maketitle

The matter-antimatter asymmetry in the Universe
is an outstanding problem of modern physics. It is expected that equal amounts of matter and antimatter
were produced in the Big Bang, yet we observe a baryon asymmetry of the order $10^{-10}$. This requires a
dynamic mechanism for baryogenesis, a prerequisite for which is violation of charge-parity (CP)
symmetry~\cite{ref:Sakharov}. While the contribution of the quark
sector is too small by several orders of
magnitude~\cite{Gavela}, leptogenesis offers a
viable alternative to generate the asymmetry~\cite{ref:Yanagida}.

Under the assumption of three-neutrino mixing and Majorana masses,
possible sources of CP violation in the lepton sector are the Dirac CP
phase $\delta_{CP}$, testable in neutrino oscillation measurements,
and the Majorana CP phases, entering lepton-number violating processes
only. Owing to the large value of the $\theta_{13}$ neutrino-mixing
angle~\cite{ref:DoubleChooz,ref:DayaBay,ref:RENO}, the $\delta_{CP}$
phase has the potential to give an important, or even dominant,
contribution to the matter-antimatter asymmetry of the
Universe~\cite{ref:Pascoli}.

To observe CP violation in neutrino oscillations, an appearance
experiment is necessary~\cite{ref:Cabibbo}. In this Letter, we
consider the $\nu_\mu \rightarrow \nu_e $ and $\bar\nu_\mu \rightarrow
\bar\nu_e $ transitions, for which the oscillation probabilities in
vacuum can be approximated
by~\cite{Cervera:2000kp}
\begin{eqnarray}
P_{\mu e} & \simeq & s_{23}^2\sin^22\theta_{13}\sin^2\Delta_{31} + c_{23}^2\sin^22\theta_{12}\sin^2\Delta_{21} \nonumber \\
& + &  \tilde{J} \cos\left( \mp \delta_{CP} -\Delta_{31} \right)\Delta_{21}\sin\Delta_{31},
\label{eq:Pemu}
\end{eqnarray}
with $s_{ij}\equiv \sin\theta_{ij}$, $c_{ij} \equiv \cos\theta_{ij}$,
$\Delta_{ij}\equiv (m_j^2-m_i^2)L/(4E_\nu)$, and $\tilde{J} \equiv c_{13}\sin
2 \theta_{13} \sin 2\theta_{12} \sin 2\theta_{23}$. Here, $E_\nu$ is
the neutrino energy and $L$ denotes the distance from the neutrino
source to the detector. Finally, the upper (lower) sign of
$\delta_{CP}$ refers to the neutrino (antineutrino) channel. Note that
in our calculations, the exact formulas including matter effects are
used instead of Eq.~\eqref{eq:Pemu}.

Unless $\delta_{CP}=0$ or $\pi$, the CP symmetry is violated and the oscillation probabilities~\eqref{eq:Pemu} are different for neutrinos and
antineutrinos. For a maximally CP-violating value of $\delta_{CP}$, a combination of ongoing oscillation experiments such as T2K~\cite{ref:T2K} and
NOvA~\cite{ref:NOvA} will probe CP violation at the $\sim 2\sigma$ confidence level. Interestingly, recent results from the
T2K experiment combined with reactor measurements for $\theta_{13}$ show some preference for maximal CP violation,
$\delta_{CP}=-90^\circ$~\cite{ref:T2K_deltaCP,Elevant:2015ska}. Nevertheless, the accuracy of T2K and NOvA in determining the value of $\delta_{CP}$ will be between 30 and 70$^\circ$ at the $1\sigma$ confidence level~\cite{Coloma:2012wq, Coloma:2012ji, deGouvea:2013onf}.

An accurate measurement of $\delta_{CP}$ has important consequences for
model building. In many flavor models, particular relations---called
sum rules---take place between the mixing angles and $\delta_{CP}$~\cite{King:2014nza,Petcov:2014laa}.
Furthermore, in models of leptogenesis with sizable flavor effects, the values of the
oscillation parameters have a strong impact on the matter-antimatter
asymmetry of the Universe~\cite{ref:Pascoli,DiBari:2010ux}.
Therefore a precise determination of $\delta_{CP}$, together with the mixing angles and light neutrino masses, can be a powerful tool to discriminate among different models and help to shed some light on some of the open questions of the Standard Model~\cite{Ballett,Meloni:2013qda}.

A precise determination of the value of $\delta_{CP}$ in future experimental
programs~\cite{ref:LBNE,ref:LBNO,ref:HyperK,Baussan:2013zcy} will
require to achieve an unprecedented accuracy, keeping systematic
uncertainties under control at the percent level. Since neutrino beams are rather broad in energy, for any given event observed at the detector the neutrino energy needs to be reconstructed from the measured kinematics of the measured particles in the final state.
Because the CP-violating phase enters the oscillation probabilities~\eqref{eq:Pemu} with a non-trivial dependence on the neutrino energy, a bias in the energy reconstruction translates into a bias in the determined $\delta_{CP}$ value, as we discuss in this Letter.

Provided the energies of all the particles produced in the event are measured, the neutrino energy in charged-current (CC) processes can be
simply determined using the calorimetric method,
\begin{equation}\label{eq:calEnergy}
E^\textrm{cal}_\nu=E_{\ell}+\sum_{i}T^N_i+\epsilon_n+\sum_{j}E_j,
\end{equation}
summing the charged lepton energy $E_{\ell}$, the kinetic energies of
the knocked-out nucleons $T^N_i$, the corresponding separation energy
$\epsilon_n$, and the total energy of any other particle produced
$E_{j}$.

Unlike the kinematic energy reconstruction, based on the charged lepton
kinematics only---typically used for quasielastic
events~\cite{ref:K2K_PRL,ref:MiniB_kappa,ref:NOMAD,ref:SciBooNE_inclusive}---the
calorimetric method~\eqref{eq:calEnergy} is applicable to any final state. This feature
is of great importance for oscillation studies performed in the
high-energy regime, where quasielastic processes give only a small
contribution to the inclusive CC cross
section~\cite{ref:Zeller_xsec}. One of the main advantages of a fine-grained time projection chamber (TPC) with respect to those detectors which only can track
the leading charged lepton is its ability to distinguish $\nu_e$ CC events from neutral-current (NC) backgrounds, even for
CC events which are either resonant or deeply inelastic. In these types
of events, multiple tracks can occur and the
identification of the leading electron track requires fine-grained
detection technique. Also, this provides the ability to measure the energy
deposited in the hadron shower.

Moreover, for an appearance experiment with neutrino energies in the GeV range, a fine-grained TPC would in principle outperform a much larger water-Cherenkov
detector thanks to the ability to reconstruct also events which are not quasielastic. However, an accurate
reconstruction of hadrons is a formidable experimental task,
especially in the case of multi-track events. One needs to keep in mind
that neutrons typically escape detection and any undetected pion leads
to an energy underestimation by at least its mass, $\sim$135 MeV. The limited
detection efficiencies for the different hadrons produced in the
events will also contribute to the missing energy budget.
Noteworthy, the number and energy distribution of hadrons
in general (and neutrons in particular) is very different for neutrino
and antineutrino CC events.

In principle, part of these effects may be alleviated by using near-detector data and by an accurate determination of the detector response to different test beams.
Nevertheless, the near and far
detectors will most generally not be identical in design or in
performance, which leads to notable uncertainties when determining the
detector capabilities. In particular, different dimensions of the near and far detectors
may result in a significantly different containment of neutral pions and neutrons.
Therefore, the effect of neutral secondaries, like neutrons, has to be corrected for by using the
detector Monte Carlo, which ultimately relies on an event generator. If
the physics model in the event generator provided an accurate
description of the underlying physics this would not present a major
problem. However, currently, event generators may not be able to provide sufficiently accurate predictions for the multiplicities and energy distributions of neutral secondaries and absorbed pions. While in the coming years, more sophisticated theoretical models can be expected to be implemented in generators, their accuracy would have to be validated by data and cannot be relied on \emph{a priori}. The spirit of this Letter is, first, to
demonstrate that there is a problem arising from missing energy and, second, to
explore the level of accuracy required in estimating this missing
energy to avoid a deleterious effect on the measurement of the CP
phase. However, we are not concerned with \emph{how} this can be practically
achieved. The detailed simulation and study of the
effect of the near detector on the determination of missing energy in
neutrino interactions is well beyond the scope of this work, and will
eventually have to be performed by the experimental collaborations. In
this work, we will instead demonstrate and quantify, from a
phenomenological point of view, how an underestimation of the missing
energy in neutrino events may affect the extraction of the value of
$\delta_{CP}$. 

To analyze the effect of realistic detection capabilities on the
energy reconstruction in a fine-grained TPC, we take into account energy
resolutions, efficiencies, and thresholds for particle detection.
For all hadrons, we set them to optimistic values, detailed in Ref.~\cite{Ankowski:2015jya}.
In our considerations, finite detector resolutions smear particle
energies according to the normal distributions centered at true energy
values. 

We make the assumption that all neutrons escape detection. It should be stressed that this assumption is rather conservative. For instance, in Ref.~\cite{DUNE-CDR} it is estimated that only a 10\% of the neutrons with energies below a GeV will escape detection in a liquid argon TPC. However, since neutrons travel some distance from the primary interaction vertex before scattering, they are problematic to associate with the neutrino event. Besides, neutrons typically deposit only part of their energy in the detector. Therefore, this assumption may need to be revisited in future, when the ongoing experimental program~\cite{Berns:2013usa, Liu:2015fiy} brings progress in the understanding of detector response to neutrons.

Very recently, several experimental collaborations have started to study in more detail the impact of systematic uncertainties on their CP violation sensitivities. In Ref.~\cite{Cherdack:2015hya}, for instance, a fast Monte Carlo was used to estimate the impact of the detector performance and nuclear effects on the neutrino energy resolution for the Deep Underground Neutrino Experiment (DUNE). Nevertheless, when computing the sensitivities to different observables, the assumptions which could affect the energy resolutions were kept fixed in the analysis. Our approach is different from the one considered in previous references: instead, in the present work we consider the possible effect in the analysis if the assumptions used to get the neutrino energy resolution were very different from expectation.

Imperfect detection capabilities induce a non-vanishing probability that an event of a true energy $E_{\nu}$ ends up being reconstructed with a different energy $E_\textrm{rec}$. We encode them in a set of migration matrices, calculated as in Ref.~\cite{Ankowski:2015jya}, using  \GENIE{} 2.8.0~\cite{ref:GENIE} with the  $\nu T$ modules package~\cite{ref:vT}. Note that our results inevitably are subject to uncertainties coming from nuclear effects. Should the argon nucleus be employed as the target, those uncertainties would not be possible to estimate, due to scarcity of reported neutrino cross sections. Therefore, in order to minimize nuclear uncertainties of our results, we consider the carbon target, for which a number of the extracted cross sections is available. This allows us to discuss the role of detector effects in an unambiguous way.

The considered experimental setup consists on a wide-band neutrino beam produced mainly from pion and kaon decays, aimed at a 40 kton detector located at a distance of $L = 1300$~km from the source. The neutrino fluxes used in this Letter correspond to the 80 GeV beam configuration from Ref.~\cite{LBNFloi}, with an assumed beam power of 1.08~MW. The background implementation follows Ref.~\cite{LBNFloi} as well. No migration matrices are used for the background events, which are always smeared according to a Gaussian with $\sigma(E_\nu) = 0.15\sqrt{E_\nu}$. As for the signal efficiencies, since the detection of neutrino and antineutrino CC events depends on the ability to observe and tag only the associated charged lepton, we use the same signal efficiencies as in Ref.~\cite{LBNFloi} (80\%), where this is taken into account. The energy of all particles produced in the event (both the charged leptons and the hadrons) are then smeared according to a Gaussian, as explained in Ref.~\cite{Ankowski:2015jya}, before reconstructing the neutrino energy. Detection thresholds and efficiencies for all hadrons are implemented as well, following Ref.~\cite{Ankowski:2015jya}. The hadron thresholds and efficiencies will affect the smearing of the events in reconstructed neutrino energy, but not the total event rates.

A total of 6 years (3 in positive horn focusing/neutrino running mode, and 3 in negative horn focusing/antineutrino running mode) are considered. Under these assumptions, the total number of events in the neutrino (antineutrino) running modes with reconstructed energies between 0.6~GeV and 6~GeV is: 740 (286) signal events, 114 (67) intrinsic beam $\nu_e$ and $\bar\nu_e$ events, 67 (33) misidentified $\nu_\mu$ and $\bar\nu_\mu$ events, and 65 (38) neutral current events. It should be noted that in the antineutrino running mode we consider both $\bar\nu_\mu \rightarrow \bar\nu_e$ and $\nu_\mu \rightarrow \nu_e$ as the signal events due to the large contribution to the signal from wrong-sign events, which are also sensitive to CP violation. In the neutrino running mode, however, only the $\nu_\mu \rightarrow \nu_e$ events are considered as part of the signal, since the wrong-sign contribution is negligible.

A modified version~\cite{Coloma:2012ji} of
\GLoBES{}~\cite{globes1,globes2} (General Long Baseline Experiment
Simulator) is used for the oscillation analysis. To determine
the confidence regions and the significance of the signal, a binned
$\chi^2$ is constructed, following the prescription of
Refs.~\cite{Coloma:2013rqa,Coloma:2013tba,Coloma:2012ji}. The bin size is set to 100~MeV in reconstructed neutrino energy. In this
work, however, no near detector is considered. Instead, we make rather aggressive assumptions
for the systematic uncertainties, and assume that the near detector
will be able to achieve these goals. Two sets of systematic
uncertainties are considered for the signal: a normalization
(bin-to-bin correlated) and a shape (bin-to-bin uncorrelated) uncertainty.
A prior at the 2\% level is considered for both of them, following
Ref.~\cite{ref:LBNE,LBNFloi}. As for the background, only a global normalization
uncertainty, at the 5\% level, is considered.

\begin{figure}
\begin{center}
	\includegraphics[width=\columnwidth]{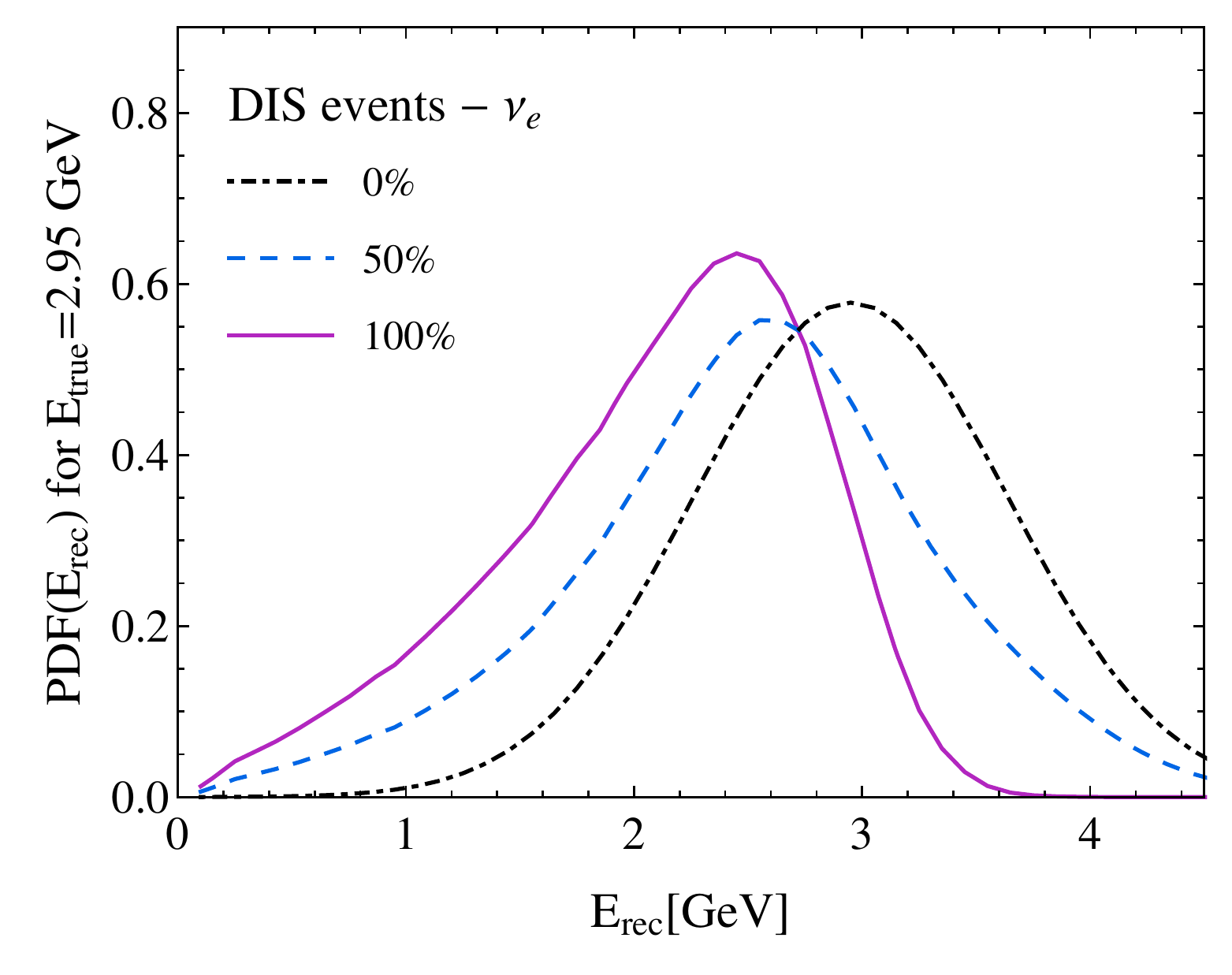}
\caption{(color online). Reconstructed energy distributions obtained for $\nu_e$ deep-inelastic scattering (DIS) events with true energy of 2.95~GeV. The distributions neglecting the shift due to the missing energy (dot-dashed line), and accounting for its 50\% (dashed line) are compared to the full calculations (solid line).
\label{fig:pdf}}
\end{center}
\end{figure}

\begin{figure}
\begin{center}
	\includegraphics[width=\columnwidth]{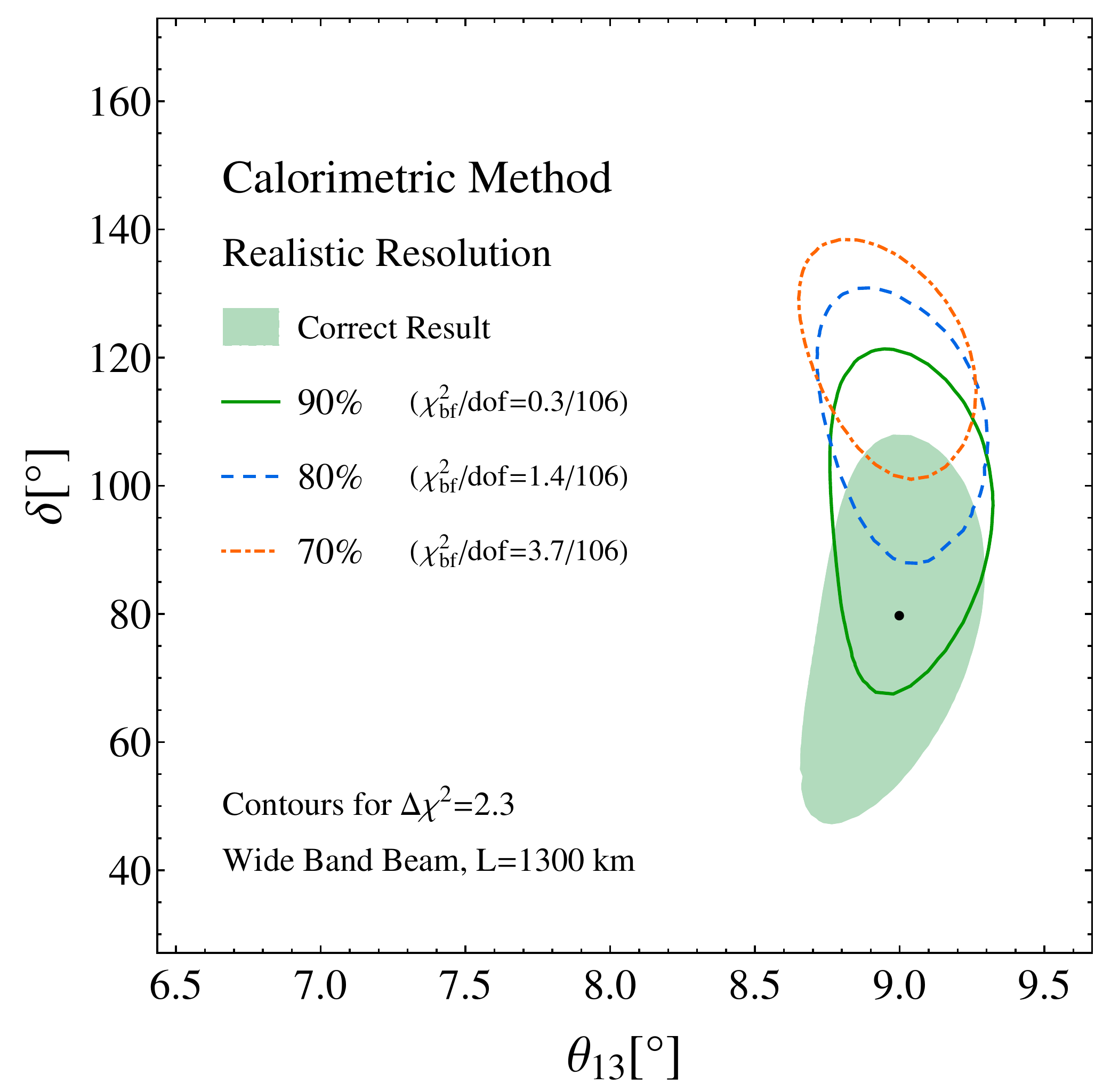}
\caption{Effect of an underestimation of the missing energy in the calorimetric energy reconstruction on the confidence regions in the
  $(\theta_{13},\,\delta)$ plane, see text for details. The true values of the oscillation parameters are indicated by the dot, and are the same for all contours shown.
\label{fig:neutrons}}
\end{center}
\end{figure}

All oscillation parameters are kept fixed in our sensitivity
calculations; the conclusions are not expected to be qualitatively affected
 if marginalization over the rest of oscillation parameters
is performed.  Since the atmospheric parameters are fixed to their
current best-fit values, and we are only interested in the $\delta_{CP}$ sensitivity, there is no need to
include $\nu_\mu$ and $\bar\nu_\mu$ disappearance channels in our
analysis. Therefore, only the results in the $\nu_e$ and $\bar\nu_e$
appearance channels are included in our fits. It should be kept in mind, though, that the measurement of the disappearance parameters may be significantly affected by either an incorrect estimate of nuclear effects and/or by an inaccurate detector calibration, as it was pointed out in Refs.~\cite{Ankowski:2015jya,Coloma:2013tba,Coloma:2013rqa}, among others. An incorrect determination of the disappearance parameters would unavoidably affect the extraction of the value of the CP-violating phase from appearance measurements.

The true event rates are obtained taking into account realistic detection capabilities
which are implemented using the migration matrices obtained from Monte Carlo events.
Therefore, the neutrino energy is not reconstructed around the true energy
but around a lower value instead, owing to the energy carried away by unobserved
particles in the final state.

The fitted event rates are smeared using a different function. In the ideal case where no particle escapes detection, the neutrino energy would be smeared according to a Gaussian distribution centered around the true neutrino energy, whose width depends on the energy smearing of the different particles observed. In our analysis, the event rates used to fit the data are smeared using a linear combination between the two cases described above: the realistic scenario where migration matrices are used, and the ideal case with a Gaussian smearing around the true energy. By varying the coefficients in this linear combination, the effective smearing function obtained can be deformed smoothly from one situation to the other. In this way, we introduce a way to manually tune the amount of missing energy in the oscillation analysis, while at the same time we account for the effect of realistic energy resolutions of the detector.

To illustrate how the energy reconstruction is affected by the missing energy, in Fig.~\ref{fig:pdf} we show an example for deep-inelastic $\nu_e$ scattering at the true energy $E_\nu=2.95$~GeV. The solid line presents the reconstructed-energy distribution calculated from the Monte Carlo simulations with all detector effects. Should no energy be missing, the distribution would be centered at the true value of the neutrino energy, as the dot-dashed curve. A common way used in the literature to parametrize the resolution in neutrino energy in oscillation experiments is by using a Gaussian function with a simple function for its standard deviation: $\sigma(E_\nu) = \alpha + \beta \sqrt{E_\nu} + \gamma E_\nu$, where $E_\nu$ is the true neutrino energy in GeV. Typical values used in phenomenological studies of liquid argon detector experiments are $\sigma(E_\nu) = 0.15\sqrt{E_\nu}$, see e.g. Refs.~\cite{Barger:2007yw,Stahl:2012exa, Adams:2013qkq, Agarwalla:2012bv}. In our case, we use the migration matrices which have been obtained from the event generator, and fit the result to a Gaussian with a width in the above form. In the case of $\nu_e$ DIS events (\textit{i.e.,} the dot-dashed curve in Fig.~\ref{fig:pdf}), the best-fit to the matrices is given by a Gaussian with standard deviation $\sigma(E_\nu) = 0.158 E_\nu +  0.13\sqrt{E_\nu}$. Finally, the dashed curve, obtained from linear interpolation between the dot-dashed and solid lines, represents an intermediate situation in which 50\% of the missing energy is accounted for: the two distributions used in the linear interpolation do have the same width, while their central value differs due to the impact of missing energy in the events. It should also be noted that, for each type of neutrino interaction considered in this work, the width of the distribution obtained when computing the migration matrices is generally different.

Based on Monte Carlo studies, the hadronic energy uncertainty in the MINOS experiment has been estimated not to exceed 8.2\%~\cite{Dytman:2008st}. However, in view of the reported difficulties with the description of nuclear effects in modern simulations~\cite{Tice:2014pgu}, our results are presented for uncertainties up to 30\%. 

The allowed confidence regions from the oscillation analysis are shown in the $(\theta_{13},\,\delta)$ plane in Fig.~\ref{fig:neutrons}. In this figure, the different contours have been obtained under different assumptions regarding the ability of the experiment to determine the missing energy involved in the events. The shaded area corresponds to the correct result, where all the missing energy in the events is perfectly estimated in the fit. The solid, dashed, and dot-dashed lines represent the results
obtained when 90\%, 80\%, and 70\% of the missing energy is correctly accounted for, respectively. Our results show that even a 20\% underestimation of the missing energy
introduces a sizable bias in the extracted $\delta_{CP}$ value.
Should an experimental analysis suffer from a 30\% underestimation of the missing energy,
it would exclude the true value of $\delta_{CP}$ at a confidence level between 2 and 3$\sigma$.

The legend in Fig.~\ref{fig:neutrons} also shows the values of the
$\chi^2$ for the best fit $(\theta_{13},\,\delta)$ points divided by
the effective number of degrees of freedom, \ie the number of data bins
minus the number of parameters extracted from the data.
In an actual experiment, this ratio would give an additional contribution
to the goodness of fit. A large enough contribution would indicate
that the model used to fit the data is not correct. Our results indicate that
such contribution would be small enough
that, from a fit to the far detector data alone, it would be virtually
impossible to realize that the energy carried away by undetected
particles is being underestimated in the fit.

In summary, we have analyzed the impact of missing energy on determination of the CP-violating phase in a long-baseline neutrino appearance experiment employing the calorimetric method of energy reconstruction.
The main source of missing energy are neutrons and other hadrons escaping detection when realistic detection capabilities are taken into account. Our results suggest that an underestimation of missing energy by as little as 20\% may result in a bias of around 1 standard deviation in the extracted value of $\delta_{CP}$. As a final remark, we would like to emphasize that, although the configuration considered in our analysis is meant to be similar to the design of DUNE, clearly much more detailed studies are necessary to draw quantitative conclusions for a specific detector setup. In particular, our treatment of the missing energy uncertainty---assumed to be equal for neutrinos and antineutrinos
and independent of the energy and interaction channel---may be regarded
as simplistic. As more realistic sensitivity estimates would require an accurate knowledge of the detector response and
inclusion of nuclear-model uncertainties, out of necessity, we leave them for future investigations within experimental collaborations.

\begin{acknowledgments}
{\it Acknowledgements.---}We are indebted to Chun-Min Jen for providing us with the events used in our
analysis, and to Omar Benhar and Davide
Meloni for numerous discussions related to the topic of this Letter.
The work of AMA and CM is supported by the National
Science Foundation under Grant No. PHY-1352106. Fermilab is operated
by the Fermi Research Alliance under contract
no. \protect{DE-AC02-07CH11359} with the U.S. Department of
Energy. PC acknowledges partial support from the European Union FP7
ITN INVISIBLES (Marie Curie Actions, PITN- GA-2011- 289442). PH is
supported by the U.S. Department of Energy under contract
\protect{DE-SC0013632} and would like to thank the Mainz Institute for
Theoretical Physics for its hospitality and partial support during
the completion of this work.
\end{acknowledgments}

\end{document}